\documentclass[showpacs,aps,PRL,twocolumn,footinbib,11]{revtex4-1}

\usepackage[utf8]{inputenc}
\usepackage{color}
\usepackage{comment}
\usepackage{lineno}
\usepackage{graphicx}
\usepackage{float} 
\usepackage[caption=false]{subfig} 
\usepackage{eqnarray,amsmath}

\def\ie{{\it i.e.}}
\def\eg{{\it e.g.}}
\def\etal{{\it et al.}}

\def\oSzo{{\bigl.^1\!S^{[1]}_0}}
\def\oSze{{\bigl.^1\!S^{[8]}_0}}
\def\tSoe{{\bigl.^3\!S^{[8]}_1}}

\def\oPoe{{\bigl.^1\!P^{[8]}_1}}
\def\tPJe{{\bigl.^3\!P^{[8]}_J}}

\def\jpsi{J/\psi}

\def\psip{\psi'}
\def\etac{\eta_c}
\def\etacp{\eta_c'}

\def\copsip{{\langle\mathcal{O}^{\psip}(\bigl.^1\! S_0^{[8]})\rangle}}
\def\coetacp{{\langle\mathcal{O}^{\etacp}(\bigl.^3\! S_1^{[8]})\rangle}}
\newcommand{\Br}{{\cal B}}

\newcommand{\eqs}[1]{\begin{equation} \begin{split} #1\end{split} \end{equation} }
\def\be{\begin{equation*}}
\def\ee{\end{equation*}}
\def\bsp#1\esp{\begin{split}#1\end{split}} 
\def\bpm{\begin{pmatrix}}
\def\epm{\end{pmatrix}}

\newcommand{\cf}[1]{{Fig.~\ref{#1}}}
\newcommand{\ct}[1]{{Table~\ref{#1}}}

\usepackage{txfonts}

 \usepackage{ifpdf}
 \ifpdf
 \usepackage[pdftex]{hyperref}
 \else
 \usepackage[hypertex]{hyperref}
 \fi

 \hypersetup{
   pdftitle={},%
   pdfauthor={},%
   pdfsubject={},%
   pdfkeywords={},%
   pdfstartview={},%
   bookmarksopen=true, breaklinks=true, debug=true, %
   colorlinks=true, linkcolor=red, citecolor=blue, urlcolor=blue
 }

\begin{document}

\title{$\etacp$ Hadroproduction at Next-to-Leading Order and its Relevance to 
$\psi'$ Production}

\author{Jean-Philippe Lansberg}
\affiliation{IPNO, Universit\'e Paris-Saclay, Univ. Paris-Sud, CNRS/IN2P3, F-91406, Orsay, France}

\author{Hua-Sheng Shao}
\affiliation{Sorbonne Universit\'es, UPMC Univ.~Paris 06, UMR 7589, LPTHE,
   F-75005 Paris, France}
\affiliation{CNRS, UMR 7589, LPTHE, F-75005 Paris, France}

\author{Hong-Fei Zhang}
\affiliation{School of Science, Chongqing University of Posts and Telecommunications, Chongqing 400065, China}

\date{\today}

\begin{abstract}
We proceed to the first study of $\etacp$ prompt hadroproduction at next-to-leading order in $\alpha_s$.
Based on heavy-quark-spin symmetry, which is systematically used in  quarkonium-production
phenomenology, we demonstrate that prompt $\etacp$ can be studied at the LHC with the existing data. We emphasize
its relevance to constrain $\psi'$ production, in the same way as the first prompt $\etac$ data at the LHC
lately strongly impacted the phenomenology of $J/\psi$ studies. 
\end{abstract}

\maketitle

\textit{Introduction} -- 
Heavy-quark-spin symmetry (HQSS), whereby soft non-perturbartive 
gluon emissions do not flip the spin of heavy quarks, is, at
the heart of all phenomenological studies (see~\cite{Lansberg:2006dh,Brambilla:2010cs,Andronic:2015wma} for reviews) of quarkonium production in nonrelativistic QCD 
(NRQCD)~\cite{Bodwin:1994jh} since more than 20 years.
This symmetry sets stringent constraints between the amplitudes 
of different non-perturbative transitions at work in quarkonium production,
encapsulated in the NRQCD Long Distance Matrix Elements (LDMEs). 
For the color-singlet (CS) transitions, one knows that the spin-triplet and spin-singlet states
have the same spatial wave function and hence their LDME are related. Prime examples
are the $\etac -
 \jpsi$ and the $\chi_{c0}-\chi_{c1}-\chi_{c2}$ systems for which the CS  LDMEs
are identical up to a mere $2J+1$ 
factor. Another well known example, central
to $J/\psi$ and $\psip$ phenomenology, 
is that of the $P$-wave LDMEs for the Color-Octet (CO) transitions, $\langle {\cal O}(\tPJe) \rangle$, 
which are also equal up to a mere $2J+1$ factor.

Besides, HQSS also very strongly constrains the polarization of the produced quarkonium.
Indeed, a heavy-quark pair produced at short distances in a given spin state 
will remain, by virtue of HQSS, in this very spin state until its hadronization in a quarkonium. Based on this, 
as early as in 1994, it has been predicted~\cite{Cho:1994ih} by Cho and Wise that a high-$P_T$ $\psi'$ 
produced by gluon fragmentation should fully inherit
its mother-gluon polarization and thus be transversely polarized [in the helicity frame]. 
The reason why current state-of-the-art NRQCD predictions~\cite{Cho:1994ih} do not follow this 
simple trend is not due to an assumed
violation of HQSS but to large QCD corrections in the short distance production of
the pair, which is not necessarily produced in a spin $\pm 1$ state, even at large $P_T$ 
as earlier expected based on LO arguments. As for now, all the NLO polarization predictions heavily 
rely on HQSS. 

Moreover, HQSS lately has attracted back the attention of many, following the 
first experimental study of $\etac$ hadroproduction at the LHC~\cite{Aaij:2014bga}. 
Indeed, the cross section measured by LHCb was found to be compatible with a negligible
contribution of CO transitions. This, in conjunction with HQSS, in turn set extremely stringent constraints
on the corresponding CO transitions at work on $J/\psi$ production~\cite{Han:2014jya,Zhang:2014ybe,Butenschoen:2014dra},
to such an extent that only one fit~\cite{Han:2014jya} currently survives
these constraints with a slight tension with the CDF polarization data~\cite{Abulencia:2007us} though.

This potentially casts  doubts onto both the 
universality of the LDMEs and the validity of the factorization conjecture of NRQCD~\cite{Nayak:2005rw,Nayak:2005rt,Nayak:2006fm}. As such, it is of paramount importance to further explore 
the $2S$ charmonium sector to see whether similar tensions occur.

In the present Letter, through a complete NLO computation, we demonstrate that 
such a first study of prompt $\etacp$ production is within the reach of
the LHCb collaboration. We even show that it is the case for a couple of 
decay channels. We also emphasize how important such
a measurement  is to advance our understanding of $\psip$ production for which,
contrary to the $\jpsi$ case, no $ep$ and $e^+e^-$ data are available to
feed in NRQCD fits.

\textit{$\etacp$ studies at hadron colliders : where do we stand?} --- 
Unlike the quarkonium spin-triplet vector states, which can decay into an easily 
detectable lepton pair, the study of the spin-singlet pseudoscalar states, such as
the $\etac$ and its radial excitation, the $\etacp$, currently relies on 
hadronic decay channels. This makes their detection a real challenge 
at hadron colliders.  As aforementioned, $\etac$  hadroproduction was
first studied by LHCb~\cite{Aaij:2014bga}. To do so, they used the 
decay channel into $p\bar p$ with a branching on the order of $1.5 \times 10^{-3}$~\cite{Olive:2016xmw}; that
is 40 times smaller than the corresponding di-muon decay branching of the 
$J/\psi$. Beside a smaller branching, such hadronic channels are much more
complicated to deal with because of the high level of the combinatorial background to be
suppressed by stringent requirements on the particle identification and by limiting the 
accessible $P_T$ range, already at the level of the trigger system.
 In its study, LHCb simultaneously reported on 
the nonprompt and prompt yields, \ie\ the $\etac$ from a $b$-hadron decay
or not. The former are significantly easier to study since the $p\bar p$ pair is displaced
with respect to the primary vertex from which emerges most of the particles constituting 
the combinatorial background.

More recently, LHCb pioneered again with the first study of $\etacp$~\cite{Aaij:2016kxn} production
in exclusive $b$ decay via the $\etacp \to p \bar p$ decay channel. From the above argument, such a {\it nonprompt} $\etacp$ detection is notably simpler than that of {\it prompt} $\etacp$ 
which we wish to motivate in the present Letter. 
Yet, this first $\etacp$ production study is very encouraging and gives us reliable 
indications on the order of magnitude of some  $\etacp$ branchings.
This is crucial in order to assess the feasibility of prompt $\etacp$ cross section 
as we wish to do here.

Indeed, with the measurements of $\frac{(\Br (B^+\to \etacp K^+)  \Br (\etacp\to p \bar p))}{(\Br (B^+\to \jpsi K^+)\Br (\jpsi\to p \bar p))}
=(1.58 \pm 0.33 \pm 0.09)\times 10^{-2}$ and knowing
that  $\Br (B^+\to \etacp K^+)=(3.4 \pm 1.8) \times 10^{-4}$, 
$\Br (B^+\to \jpsi K^+)=(1.026 \pm 0.031) \times 10^{-3}$ and
$\Br (\jpsi\to p \bar p)=(2.120 \pm 0.029 ) \times 10^{-3}$~\cite{Olive:2016xmw}, 
one can derive that $\Br (\etacp\to p \bar p) = (1.0 \pm 0.5) \times 10^{-4}$, 
which we will use instead of the current PDG upper limit of $10^{-4}$.

The ratio $\Br (\etacp\to p \bar p)/\Br (\etac\to p \bar p)$ is thus likely as high as 
10~\%, which gives us great confidence that prompt $\etacp$ studies are within 
the reach of the LHCb detector with data on tape or to be recorded soon. Moreover, it is 
reasonable to suppose that the corresponding ratios for the
heavier final states $\Lambda \bar \Lambda$, $\Lambda^* \bar \Lambda^*$,  $\Xi  \bar \Xi$ would
be on the same order than for the $\etac$, if not larger, since, in 
general, the phase space is relatively larger for $\etacp$ compared to $\etac$.
Given that $\Br (\etac\to \Lambda \bar \Lambda)= (1.09\pm 0.24) \times 10^{-3}$ 
and $\Br (\etac\to \Xi  \bar \Xi)= (8.9\pm 2.7) \times 10^{-4}$, one can 
infer that the corresponding branchings for the $\etacp$ are
expected to also be on the order of $10^{-4}$. As discussed in~\cite{Barsuk:2012ic}, 
the latter channels, in spite of the request to find 4(6) tracks, are also promising owing
to the presence of 2(4) secondary vertices which drastically reduces the combinatorial 
background. 

Beside the baryon-antibaryon channels, the $\phi\phi$ decay channel is also of interest
given the large $\phi$ branching to charged $K$ pairs resulting in two narrow and clean 
peaks~\cite{Barsuk:2012ic}. This channel is indeed the first via which $\etacp$ production
in inclusive $b$ decay was observed by LHCb~\cite{Aaij:2017tzn} and they measured
$\frac{(\Br (b\to \etacp X)  \Br (\etacp\to \phi\phi))}{(\Br (b\to \etac X)\Br (\etac\to \phi\phi))}
=(4.0 \pm 1.1 \pm 0.4)\times 10^{-2}$. Along with 
$\Br (\etac \to \phi \phi) = (1.75 \pm 0.20) \times 10^{-3}$~\cite{Olive:2016xmw}[We take
the "fit" PDG value, rather than the "average" PDG which is much less precise.]
and assuming that the $2S/1S$ ratio of the partial $b$ width for the
spin-singlet states ($\eta_c(nS)$) is similar to that of the spin-triplet states ($\psi(nS)$)
[and approximately accounting for the different feed-down structure],
we  expect $\Br (\etacp\to \phi\phi)$ to be 6-7 times smaller, that is on the order
of $2.5 \times 10^{-4}$. To phrase it differently, the di-$\phi$ channel is another serious
alternative.

Finally, the  $K \bar K \pi$ channel with a branching of about 2~\%~\cite{Olive:2016xmw}
or the $K^+ K^- \pi^+ \pi^-$ one, which should be of a similar magnitude, could be used to 
increase the reach to large $P_T$ where
the combinatorial background is getting less problematic and the statistics
small. As we shall explain below, the large $P_T$ region is probably the most interesting.
The di-photon channel ($\Br (\etacp \to \gamma\gamma) = 
(1.9 \pm 1.2) \times 10^{-4}$~\cite{Olive:2016xmw}) is a further option to be 
explored. Yet another method would be to look for $\etac$ or $\etacp$ in the 
already recorded $J/\psi$ sample as suggested in~\cite{Lansberg:2013qka}. 
This would allow one to bypass most of the triggering constraints. 
\ct{tab:etacp_branching} summarizes the aforementioned channels and their status.

\begin{table}[hbt!]
\begin{tabular}{c|c|c}
Channel & $\etacp$ partial width & status \\
\hline
$p \bar p$ & $(1.0\pm 0.5) \times 10^{-4}$ & measured \\ 
$\gamma \gamma$ & $(1.9 \pm 1.2) \times 10^{-4}$ & measured  \\
$K \bar K \pi$ & $(1.9 \pm 1.2) \times 10^{-2}$ & measured\\
$\phi \phi$ & $(1 \div 4) \times 10^{-4}$ & extrapolated  \\

$\Lambda \bar \Lambda$ & ${\cal O}(10^{-4})$ & estimated \\
$\Xi  \bar \Xi$ & ${\cal O}(10^{-4})$ & estimated \\
\end{tabular}
\caption{Summary of tractable $\etacp$ decay channels in hadroprodution studies.}
\label{tab:etacp_branching} 
\end{table}

\textit{Theoretical framework ---} 
Within NRQCD, one factorizes any differential cross section to produce a 
quarkonium into calculable coefficients related to the production at short distances of
the heavy-quark pair in different Fock states and
the LDMEs encoding the non-perturbative transitions between these states and the final-state 
quarkonium. Generally, the partonic quarkonium-production cross section reads
\eqs{
\sigma(\mathcal{Q}+X)=\sum_{n}{\hat{\sigma}(Q\bar{Q}[n]+X)\langle O^{\mathcal{Q}}(n)\rangle},
\label{eq:NRQCD}}
where $n$ labels the different Fock states of the $Q\bar{Q}$ pair which is produced in the partonic
scattering. Correspondingly, the $\hat{\sigma}(Q\bar{Q}[n]+X)$ are the short-distance coefficients (SDC),
which are perturbatively calculable using usual Feynman graphs.  
$\langle O^{\mathcal{Q}}(n)\rangle$ denotes the nonperturbative --but universal-- NRQCD LDMEs. 
The hadronic cross section is then obtained by folding $\sigma(\mathcal{Q}+X)$ with the parton distribution functions (PDFs) as usually done in collinear factorization.

The predictive power of NRQCD relies on the truncation of the sum over only a couple of Fock states $n$, 
namely those whose LDMEs are expected to be the least suppressed in powers of the velocity, $v$,
of the heavy quarks in the rest frame of the pair. Indeed, NRQCD relies on a double expansion
in both $\alpha_s$ and $v$. As a case in point, the direct production of $\etacp$ receives contributions from four Fock states up to $\mathcal{O}(v^4)$, i.e. $\oSzo,\oSze,\tSoe,\oPoe$. 

At hadron colliders, the most common analyzed observable remains the $P^{\cal Q}_T$-differential cross section. 
Explicit NLO $(\alpha_s^4)$ computations of the SDCs in the LHC kinematics show that the sole $\oSzo$ and $\tSoe$ states 
are relevant to predict it -- unless unrealistically large LDMEs are allowed for the other transitions. In addition, we note that for $C=+1$ states (unlike the $C=-1$ states like the $J/\psi$) 
the leading $P^{\cal Q}_T$ topologies already appear at $\mathcal{O}(\alpha_s^4)$, i.e. at NLO in $\alpha_s$.
NNLO corrections should thus be mild.  
  
As announced, we have performed our analysis at NLO in $\alpha_s$. To do so, we relied 
on the FDC framework~\cite{Wang:2004du,Gong:2014qya} which generates the Born, real-emission
and virtual contributions, ensures the finiteness of their sum, performs the partonic-phase-space integration and that
over the PDFs. We also note that the SDCs for $\etac$ and $\etacp$ are 
equal up to relativistic corrections, suppressed at least as $v^2$.

We then used the LDMEs obtained from the $\psip$ NLO 
fits~\cite{Shao:2014yta,Gong:2012ug,Bodwin:2015iua} and, by virtue of HQSS,
set $\coetacp=\copsip$ (see \ct{tab:LDME}). For the CS channel, we take $|R(0)|^2=0.53$~GeV$^3$ for
the radial wave function at the origin.
Contrarily to the $J/\psi$ case, these fits cannot rely on $e^+e^-$ nor $ep$ data. Polarization
data are also more limited which explains that only three NLO fits exist in the literature.

\begin{table}[hbt!]
\begin{footnotesize}
\hspace{-0.5cm}
\begin{tabular}{c|c|c|c}
\hline
\rule{0pt}{3ex}
& Shao \etal~\cite{Shao:2014yta} & 
 Gong \etal~\cite{Gong:2012ug} & Bodwin \etal~\cite{Bodwin:2015iua}\\[1mm] 
\hline
\parbox{2.cm}{$\coetacp$ [$10^{-2}$ GeV$^3$]}& $[0,3.82]$  
  & $[-0.881,0.857]$ 
  & $[2.35,3.93]$\\   
\hline
\end{tabular}
\caption{Allowed range of the LDME $\coetacp$ from 
3 existing NLO fits of $\copsip$.}
\label{tab:LDME} 
\end{footnotesize}
\end{table}

As for the used parameters, we have set $m_c$ to 1.5 GeV and computed the cross
section using the renormalization and factorization scale values, $\mu_R$ and $\mu_F$,
within the pairs
$(\mu_R,\mu_F)=\mu_0 \times (1,1;0.5,0.5;2,2;0.5,1;1,0.5;1,2;2,1)$ with $\mu_0^2=4m^2_c+P_T^2$.
The uncertainty on $m_c$ is not shown as it is necessarily strongly correlated --unlike the scale uncertainties-- to that in the extraction of the LDME with $\psip$ data. We have used the default PDF sets in FDC, namely CTEQ6M (CTEQ6L1) for the NLO (LO) curves which are the ones so far used NLO LDME fits.

We note that despite this large spread of the $\copsip$,
the 3 fits all reproduce the $\psip$ LHC data. The variation in the size of 
$\copsip$ is compensated by that
of $\langle {\cal O}^{\psip}(\tSoe) \rangle$ and $\langle {\cal O}^{\psip} (\tPJe) \rangle$ --a well known phenomenon for the $J/\psi$ which is however constrained by polarization and $ep$ data.
In practice, the existing $\psip$ data do not suffice to lift the degeneracy between 
these LDMEs driving the $P_T$ behavior of the $\psip$ differential cross section.
As we know, the situation is completely different for the $\eta_c$ where 
$\coetacp$ is essentially the only relevant CO transition.
The same of course happens for the $\etacp$ differential cross section as we will show now.

\textit{Results and discussion} ---  
\cf{fig:CO_vs_CS-LO_vs_NLO} shows our (N)LO cross sections for the direct $\etacp$ yield 
at $\sqrt{s}=13$~TeV in the rapidity acceptance of the LHCb detector. On purpose, we
have not combined the CS and the CO contributions since the latter is essentially unconstrained. 
The displayed CS curve can be seen as a lower value~\footnote{Only a negative CO LDME can 
yield to a smaller cross section. This is only allowed by the Gong \etal\ fit and
in general is known to yield to negative --and unphysical-- cross sections for some associated-production processes, 
see \eg\ \cite{Li:2014ava}.} fit and the CO one --with the chosen LDME value, $\coetacp=0.0382$ GeV$^3$--
is a likely upper limit. In addition to the size of the cross section, on which we elaborate further
below, we note that of the impact of the NLO corrections differs in each case. In the CO case, one
observes a classical reduction of the theoretical uncertainties along the lines of a good
convergence of the perturbative series. In practice the $K$ factor happens to be slightly below unity at large $P_T$. 
In the CS case, the main effect is the expected appearance of new leading $P_T$ topologies which results
in a $K$ factor increasing with $P_T$ (see \cite{Artoisenet:2008fc,Lansberg:2008gk,Lansberg:2009db,Gong:2012ah,Lansberg:2013qka} for analogous cases). Since the leading-$P_T$ CS contributions eventually consist in $\alpha_s^4$
Born topologies, the $\mu_R$ sensitivity also naturally increases in the absence
of the corresponding virtual corrections to these topologies which only appear at $\alpha_s^5$. 

\begin{figure}[hbt!]
\centering
\includegraphics[width=1.0\columnwidth,draft=false]{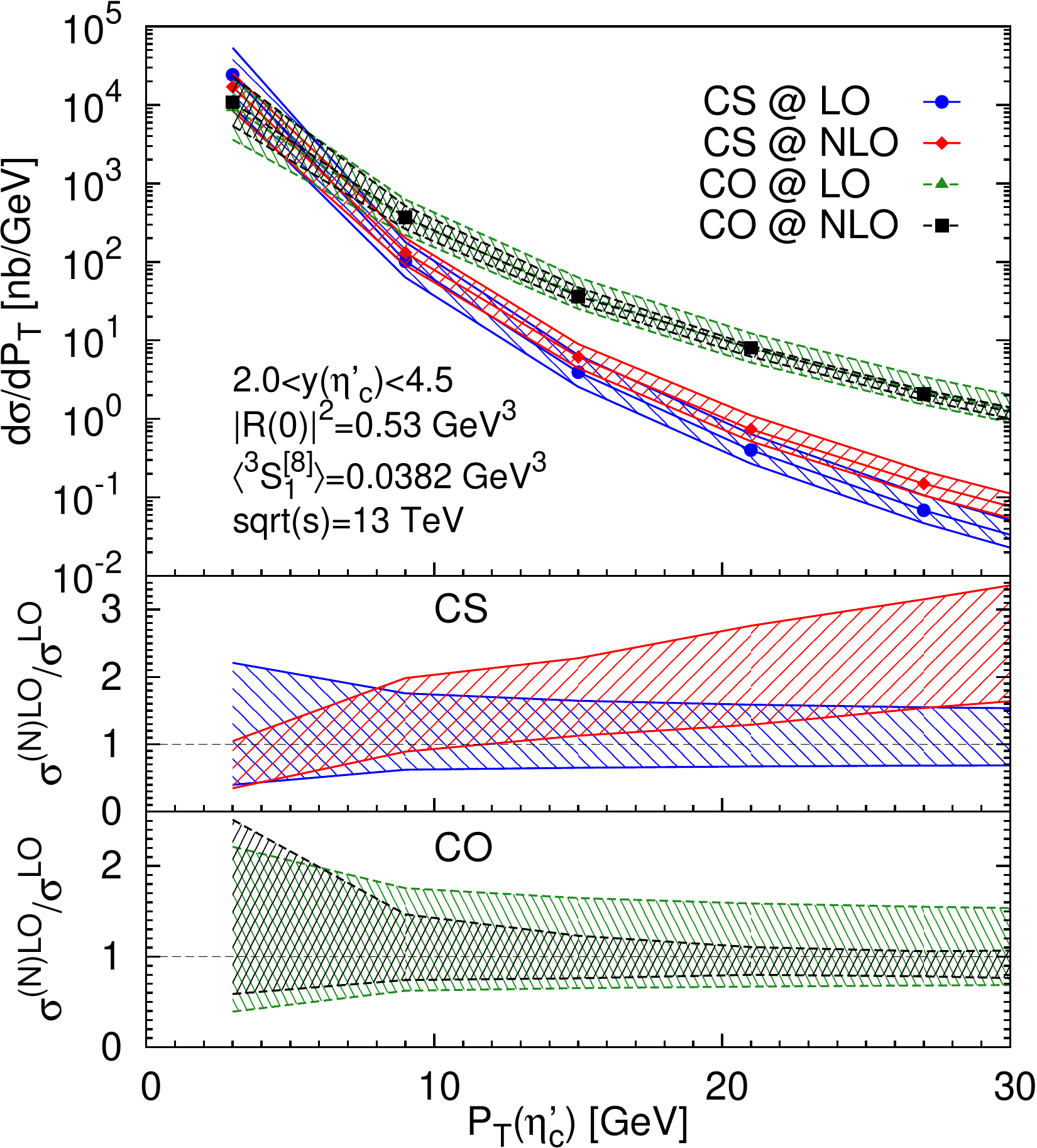}
\caption{Differential-$P_T$ cross section for $\etacp$ production: comparison between
the LO and NLO predictions for both the CS and CO channels. The latter is plotted using the upper
value of the Shao \etal\ fit.}\label{fig:CO_vs_CS-LO_vs_NLO}
\end{figure}

\begin{figure*}[hbt!]
\centering
\subfloat{\includegraphics[width=0.33\textwidth,draft=false]{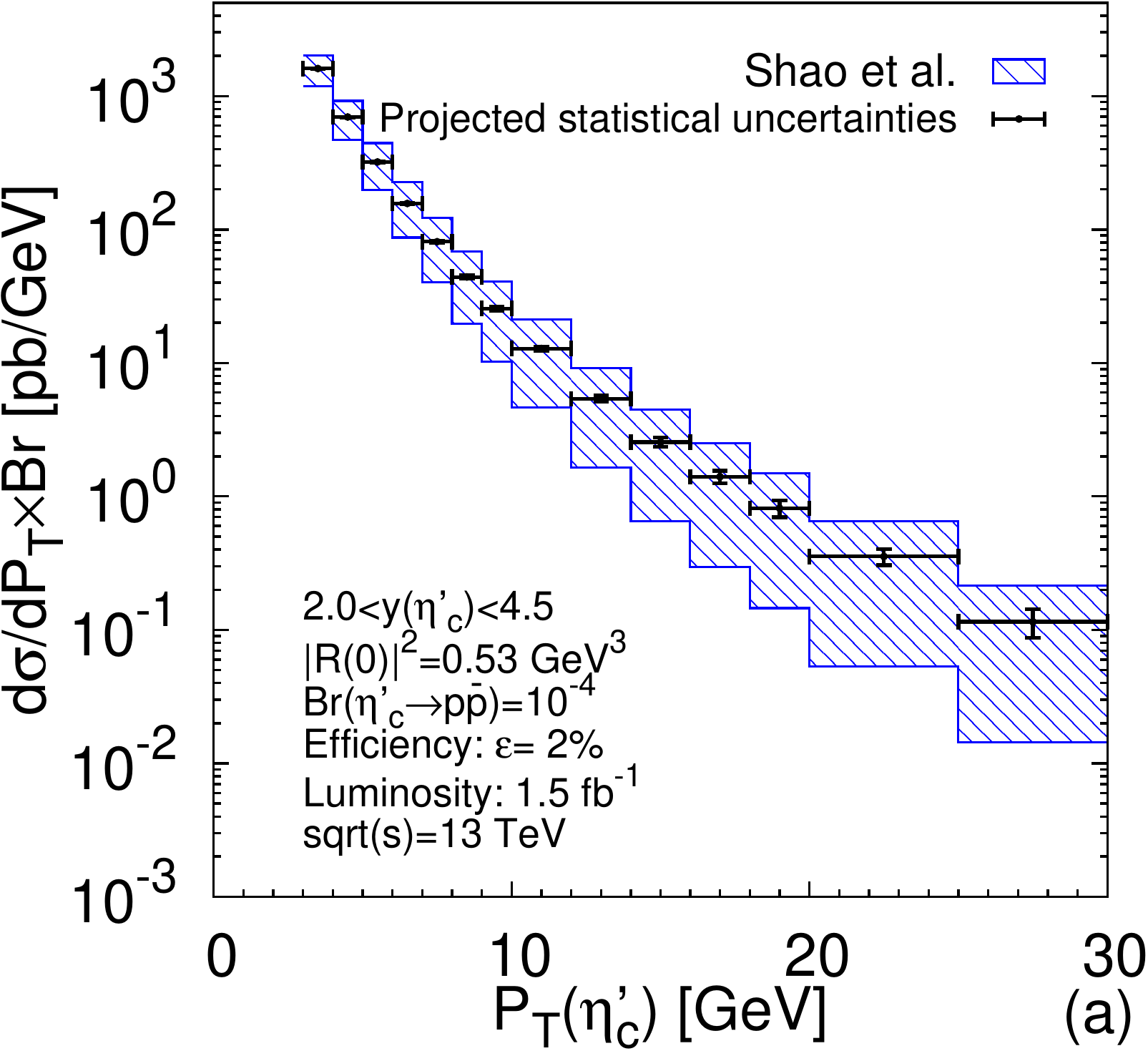}\label{fig:Shao}}
\subfloat{\includegraphics[width=0.33\textwidth,draft=false]{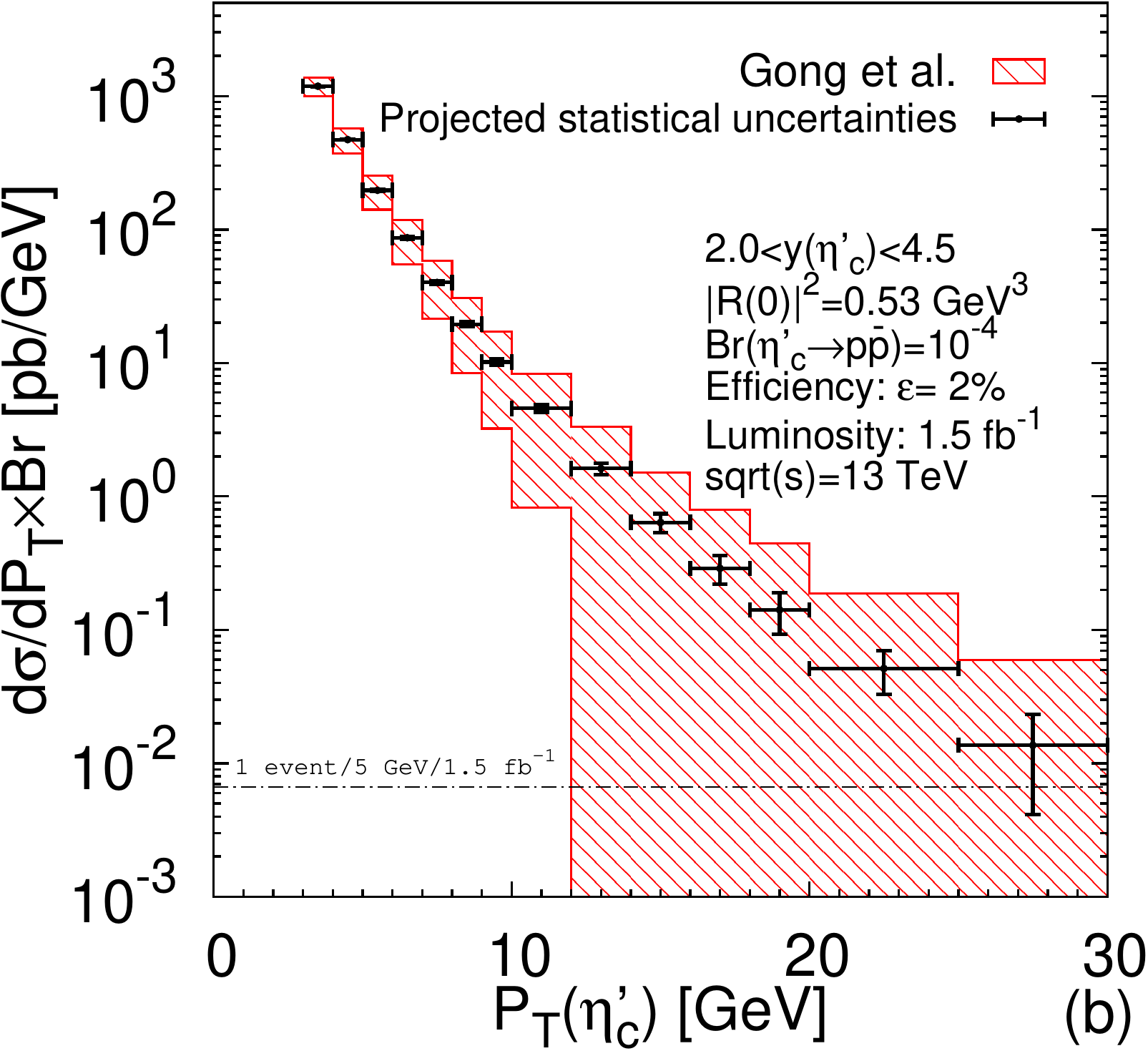}\label{fig:Gong}}
\subfloat{\includegraphics[width=0.33\textwidth,draft=false]{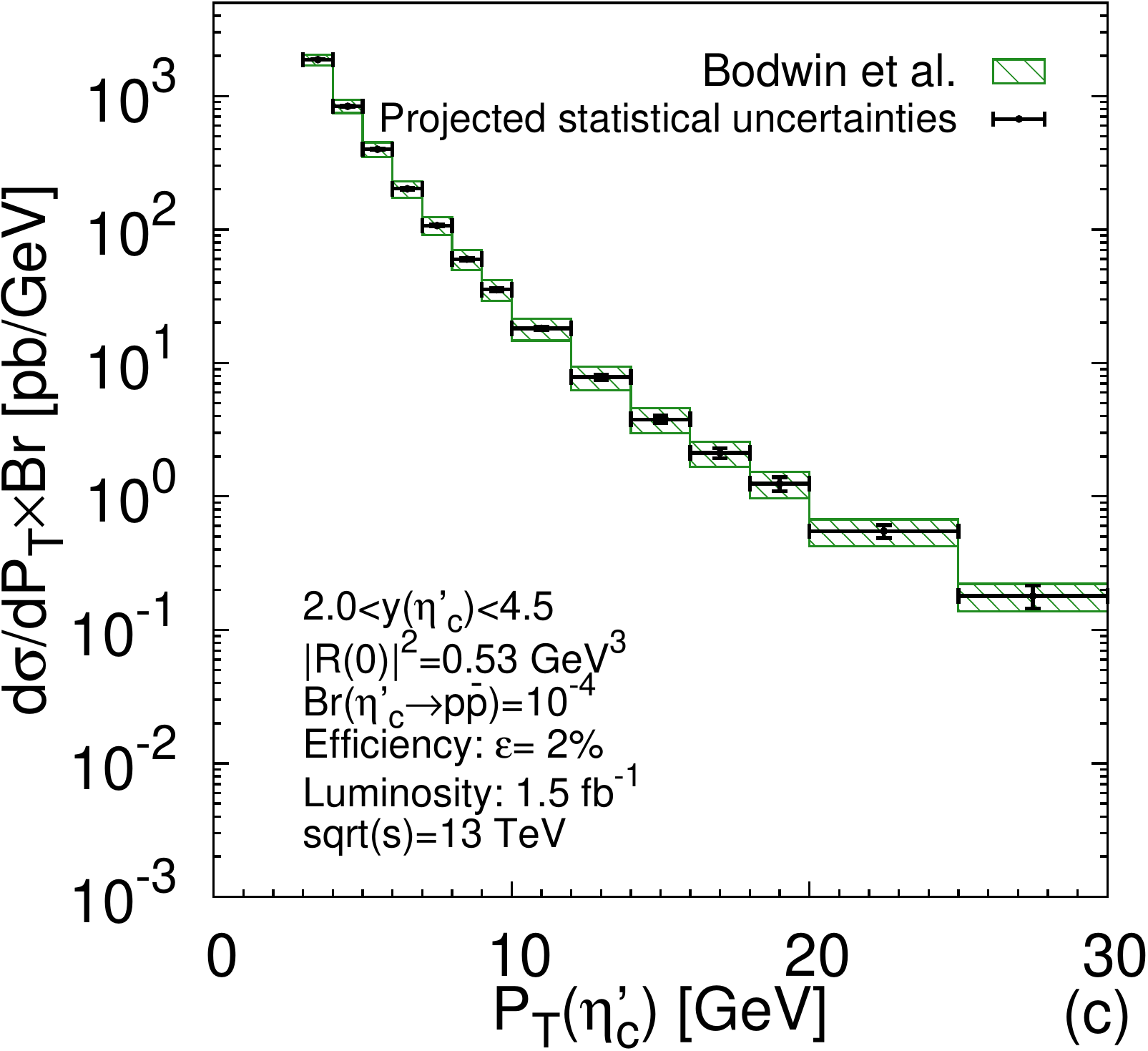}\label{fig:Bodwin}}
\caption{Differential-$P_T$ cross section for $\etacp$ production times ${\cal B}(\etacp \to p \bar p)$ 
for the 3 ranges of $\coetacp$ (a,b,c) along the 
projected statistical uncertainties using the central theoretical values in each cases, with
an assumed efficiency of 2\% and the luminosity collected so far by LHCb at $\sqrt{s}=13$~TeV, 1.5~fb$^{-1}$.
\label{fig:3-CO}}
\end{figure*}

In addition, we note that the large difference between the CS and CO curves reinforce our confidence
in the discriminating power of forthcoming $\etacp$ data, even with a moderate precision, especially above $P_T=10$ GeV where 
the $\tSoe$ channel should dominate. 

In fact, data already on tape or to be recorded very soon most probably offer enough precision in this region.
\cf{fig:3-CO} shows the projected statistical uncertainties expected
with the central value of theoretical range for each LDME fit. It has been 
derived with a detection efficiency of 2~\%~\footnote{That is a similar efficiency as
for the $\eta_c$ derived from \eg\ 
$\varepsilon(\etac) / \varepsilon(\jpsi) \sim ~ N_{\rm obs}(\etac)/N_{\rm obs}(\jpsi) \times 
\sigma(\jpsi)/\sigma(\etac) \times {\cal B}(\jpsi \to p\bar p)/{\cal B}(\etac \to p\bar p)$ where
$N_{\rm obs}$ and $\sigma$ are the numbers of candidates in the signal peaks and cross section
quoted in~\cite{Aaij:2014bga}. Assuming similar detection efficiencies for $\etac$ and $\etacp$ is a reasonable working hypothesis given that the $\etacp$ trigger efficiency may be little higher with slightly more energetic decay products
whereas its signal-over-background ratio is likely similar (smaller signal but smaller background).} in the $p\bar p$ decay channel
and with a luminosity --already collected at $\sqrt{s}=13$~TeV-- of 1.5 fb$^{-1}$.
Clearly, the $\etacp$ cross section should be measurable up to $P_T=20$ GeV with a decent precision, and significantly further if other channels can be used as already done for the nonprompt sample.

Going further, one sees that the expected cross sections significantly differ depending on the
$\psip$ fit used, compare  \eg\ \cf{fig:Gong} and \cf{fig:Bodwin}. 
 It is clear that forthcoming $\etacp$ data will significant narrow down the allowed range 
for $\coetacp$, and thus of $\copsip$. In fact, if the measured cross section
coincides with that predicted based on the Bodwin \etal\ $\psip$ fit, it would virtually 
exclude the Gong \etal\ $\psip$ fit and reduce by an order of magnitude the range of 
the Shao \etal\ $\psip$ fit. Conversely, since the CS cross section [with a vanishing CO LDME] 
nearly coincides with the central value of the Gong \etal\ band, one also concludes 
that a cross section compatible with the CS contribution alone --as for the $\etac$~\cite{Aaij:2014bga}-- 
would have the statistical power to exclude the Bodwin \etal\ fit as well as to significantly
reduce the range of the Shao \etal\ fit. To phrase it differently, a measurement 
compatible with with the CS contribution alone directly sets an upper limit 
on $\copsip$.

\textit{Conclusions.---}
We have performed the first --and complete-- one-loop analysis
of $\etacp$ hadroproduction. We have focused on the LHC case, in particular on 
the potential of the LHCb experiment to measure its cross section in a $P_T$ range
which would severely constrain --by virtue of HQSS-- $\psip$ theoretical predictions. 

After discussing the usable $\etacp$ decay channels, we have demonstrated
that such studies are well within the reach of LHCb already with data on tape with a
channel which they already studied for the nonprompt yield. 
Given the smaller $\psip$ data set compared to the $\jpsi$ ones and the quasi absence
of feed down effects, we expect that the impact of such a measurement will surpass that
of the first LHCb study of $\etac$ production at 7 and 8 TeV and to impact the field at large.

\textit{Acknowledgements.---}
We are indebted to S. Barsuk and A. Usachov for essential comments on our study.
The work of J.P.L. is supported in part by the French IN2P3--CNRS via the LIA FCPPL 
(Quarkonium4AFTER) and the project TMD@NLO. 
The works of H.-S.S was supported by the ILP Labex (ANR-11-IDEX-0004-02, ANR-10-LABX-63).
H.F.Z is supported by the National Natural Science Foundation of China (grant No. 11405268).

\bibliographystyle{utphys}

\bibliography{etac2S-011117.bib}

\end{document}